\documentclass[%
 reprint,twocolumn,
superscriptaddress,
 amsmath,amssymb,
 aps,
]{revtex4-1}

\usepackage{graphics}      
\usepackage{graphicx,float}      
\usepackage{longtable}     
\usepackage{url}           
\usepackage{bm}            
\usepackage[usenames]{color}
\usepackage[dvipsnames]{xcolor}
\usepackage{cancel}
\usepackage{soul}
\usepackage{amsmath,amssymb}
\usepackage{epstopdf}
\usepackage{ulem}
\definecolor{lightbrown}{RGB}{210, 180, 140}

\begin{document}


\title{Characterizing some dynamical states in swarmalators system using recurrence analysis}

\author{Delors A. Jiofack}
\affiliation{Department of Physics, School of Philosophy, Sciences and Letters, University of São Paulo, 14040-901 Ribeirão Preto, SP, Brazil.}

\author{Zidane Choffo}
\affiliation{MoCLiS Research Group, Dschang, Cameroon.}
\affiliation{Research Unit Condensed Matter, Electronics and Signal Processing, University of Dschang, P.O Box 67 Dschang, Cameroon.}

\author{Carmel T. Lambu}
\affiliation{MoCLiS Research Group, Dschang, Cameroon.}
\affiliation{Research Unit Condensed Matter, Electronics and Signal Processing, University of Dschang, P.O Box 67 Dschang, Cameroon.}

\author{Sandrine J. Mabekou}
\affiliation{MoCLiS Research Group, Dschang, Cameroon.}
\affiliation{Mechanics and Modeling of Physical Systems Research Unit (UR-2MSP), University of Dschang, P.O Box 67 Dschang, Cameroon.}

\author{Patrick Louodop}
\affiliation{MoCLiS Research Group, Dschang, Cameroon.}
\affiliation{Research Unit Condensed Matter, Electronics and Signal Processing, University of Dschang, P.O Box 67 Dschang, Cameroon.}
\affiliation{ICTP South American Institute for Fundamental Research, S\~ao Paulo State University (UNESP), Instituto de F\'{i}sica Te\'{o}rica, Rua Dr. Bento Teobaldo Ferraz 271,
Bloco II, Barra Funda, 01140-070 S\~ao Paulo, Brazil.}
\affiliation{Potsdam Institute for Climate Impact Research (PIK)
Member of the Leibniz Association P.O. Box 60 12 03 D-14412 Potsdam Germany.}

\author{Fernando F. Ferreira}
\affiliation{Department of Physics, School of Philosophy, Sciences and Letters, University of São Paulo, 14040-901 Ribeirão Preto, SP, Brazil.}
\affiliation{School of Arts, Sciences and Humanities, University of São Paulo, 03828-000 São Paulo, Brazil.}

\author{Norbert Marwan}
\affiliation{Potsdam Institute for Climate Impact Research (PIK)
Member of the Leibniz Association P.O. Box 60 12 03 D-14412 Potsdam Germany.}

\date{\today}
\begin{abstract}
Chimera or chimera-like states arise in a wide variety of networks and their identification remains challenging particularly when mobility prevents index-based ordering of the nodes. In this work, we propose a recurrence analysis based method to identify and characterize chimera states in two distinct dynamical frameworks: a network of chaotic Colpitts oscillators and a system of swarmalators where delayed interactions induce chimera-like dynamics named boiling state. The suggested strategy is based on the joint recurrence plots and entropy-based measures, to capture the spatio-temporal organization. This approach enables a clear discrimination between complete synchronization, quasi-synchronization and disordered regimes, even when conventional order parameters yield ambiguous results. Furthermore, we introduce the degree of independence, which estimates the proportion of dynamically completely independent nodes in the system. This measure provides a robust characterization of transitions between collective states.

\end{abstract}

\maketitle
\section{Introduction}\label{intro}

Complex dynamical systems consisting of a large number of interacting oscillators exhibit a wide variety of collective behaviors, ranging from total desynchronization to complete synchronization. Generally between these two extremes lie various dynamics such as chimera states \cite{chim1,chim2}, hybrid regimes characterized by the coexistence of coherent and incoherent subsets within the same network \cite{chim3,chim4,chim5}. Since its discovery by Kuramoto and Battogtokh \cite{Kuramoto2002}, this state has attracted considerable interest \cite{Abrams2004,Abdoulaye2025,Gopal2014}, particularly for their potential implications in neuroscience \cite{neuro1,neuro2}, animal population dynamics \cite{anim1,anim2,anim3}, and distributed technological systems \cite{techno1,techno2,techno3,techno4}. In 2017, it was found by G. Petrungaro et al. \cite{mochi} that, mobility may induce a persistent chimera state while they were studying mobile systems meaning coupled nodes for which positions vary with time. The same year, a recent form of mobile systems was proposed and named swarmalators \cite{swarm0}. This model seems more adapted to tackle the dynamics of many biological mobile systems such as schools of fishes\cite{swarmob}.

Recently a chimera-like dynamic was found in swarmalators systems and named boiling state showing a central region where the particles were phase synchronized, steady or pulsating and surrounded by active and non-synchronous ones \cite{Blum2024,lambu}. Thus one can investigate the chimera behavior according to space (coherent region: central steady or pulsating nodes, incoherent region: active nodes) and according to the internal dynamics (coherent region : central nodes that synchronized, incoherent region : nodes at the border that are non-synchronous) \cite{Blum2024, lambu}. However, strictly speaking chimera states were easily confirmed by the perfect difference between both coherent and incoherent domains through the spatiotemporal plots of the network within which nodes are identified by their indexes. This perfect domain's separation is due to the immobility of the nodes. This is not the case of mobile systems like swarmalators whose movements interchange particles positions leading to a noisy-like spatiotemporal patterns. Thus, one need to reorganize elements before this specific representation. Also, in these last studies \cite{Blum2024, lambu}, a new state was found due to the insertion of delay in internal variables named the quasi static sync state or ring static state where particles are distributed in concentric rings such that those of the same ring are synchronous but phase-locked with the elements of other rings. This dynamics is difficult to separate from the sync state using the Kuramoto order parameter, as in both cases the order parameter's values are very close to one ($R \rightarrow 1$).

In this work, we focus on characterizing some dynamical behaviors observed in swarmalators with delayed interactions \cite{Blum2024,lambu}. Two principal behaviors are considered, that are the boiling state and the ring static state. Since the boiling state is a chimera-like motion, we will investigate the chimera state of a recently developed Colpitts oscillator \cite{Colpitts2,Kennedy1994,Kountchou2020, Carmago2024} in nearest neighbors network topology \cite{network1,network2}. The objective is to compare the results of the investigations using recurrence analysis and the strength of incoherence $(SI)$ \cite{Gopal2014}  in a network of coupled Colpitts oscillators. The insights gained from this comparison will be used to assess the applicability of recurrence analysis to swarmalator systems, where the continuous rearrangement of oscillators makes index-based measures such as $SI$ difficult to apply. \cite{CH23,CH24,CH25,CH26,CH27,CH28}. Consequently, recurrence analysis may provide a more robust framework for the identification and characterization of chimera states in mobile and dynamically reorganizing systems.  

With this in mind, we investigate the recurrence analysis parameters that could help to identify signatures of coherence and incoherence without requiring prior reorganization of the nodes. In addition, we introduce the degree of independence, which allows us to evaluate the proportion of nodes that are completely independent, i.e., uncorrelated with the rest of the system's elements. Such a measure could pave the way for applications in the study of complex biological systems, particularly for assessing the health status of patients with certain neurological failures \cite{bio1,bio2}.
 
This work has five main parts. After the introduction in Sec.\ref{intro}, We start our study by presenting both Colpitts oscillators based network and the swarmalators system that are the targets of our investigations in Sec.\ref{models}. Sec.\ref{tools} is devoted to the development of the analysis tools allowing the characterization of the obtained dynamics and their interpretation; meanwhile Sec.\ref{results} presents numerical results for both Colpitts and swarmalators systems followed by conclusion and some remarks in Sec.\ref{conclusion}.

\section{Models}\label{models}

\subsection{Colpitts Oscillator based Network}

The Colpitts model used in this work is proposed in \cite{Kountchou2020} where the nonlinear function is generated by a voltage comparator. It is a classic electronic circuit whose normalized equations describe the evolution of the capacitors voltages and current flowing through the coil. In the case of a network of $ N $ coupled oscillators, the equations for each node $ i $ are written as:
\begin{equation}
\label{col}
\left\{ {\begin{array}{*{20}{l}}
{{{\dot x}_i} =  - {x_i} - {z_i} + {v_{cc}} {\rm{sign}}({V_{ref}} - {y_i}) + {g_x},}\\
{{{\dot y}_i} =  - {y_i} + {z_i} + {v_{cc}} {\rm{sign}}({V_{ref}} + {x_i}) + {g_y},}\\
{{{\dot z}_i} = \beta ({x_i} - {y_i}).}
\end{array}} \right.
\end{equation}
With $V_{ref}$, $v_{cc}$ and $\beta$ the system's parameters. The coupling functions $g_k$ are defined as:
\begin{align}
{g_k} = r\sum\limits_{j = i - p}^{i + p} {\left( {{k_j} - {k_i}} \right)}, k = x,y
\end{align}
$r$ represents the coupling coefficient and $p$ is the number of close neighbors considered.  The following parameters values $V_{ref}= 0.3$, $v_{cc}= 15$ and $\beta= 5$ {cause a single system to exhibit chaotic behavior as shown in Fig.\ref{port} and are used throughout this work.
\begin{figure}[htbp]
    \centering
    \includegraphics[width=5cm,height=4cm]{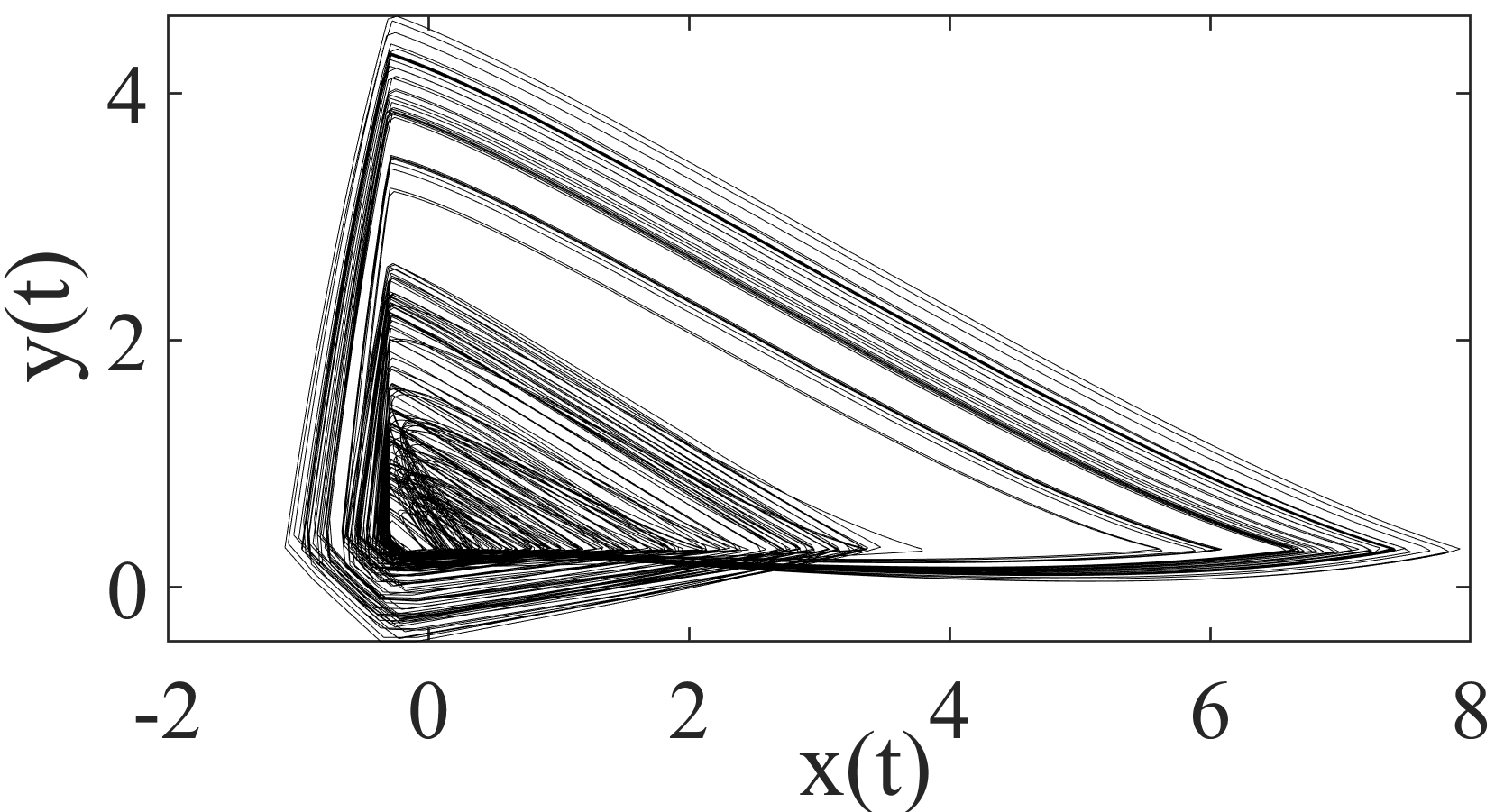}
    \caption{Chaotic dynamics of the Colpitts oscillator for which model is given by Eq.\ref{col} for the parameters $V_{ref}= 0.3$, $v_{cc}= 15$ and $\beta= 5$.}
    \label{port}
\end{figure}

\subsection{Delay swarmalators Model}

Introduced by O'Keeffe et al. \cite{swarm0}, the swarmalators model combines two types of interactions that are: a spatial interaction (attraction/repulsion) which determines the position of agents and phases interactions that lead particles internal states to different behaviors. However a third interaction can be stated describing how their internal dynamics influences the spatial motions through the coupling $J$ \cite{swarm0,Blum2024}. Each agent $ i $ is described by a spatial position ${X_i{\rm{(t)}}(x_i(t),y_i(t))}$ and a phase ${\theta _i}{\rm{(t)}} \in [0,2\pi ]{\rm{ }}$. The delayed model includes a constant delay $\tau$ in phases interactions, representing transmission or reaction time, which enriches the collective dynamics. The typical equations are written as:
\begin{equation}
\label{eq1}
\begin{array}{l}
{{\dot X}_i} = {v_i} + \frac{1}{N}\sum\limits_{j \ne i}^N {\frac{{{X_j} - {X_i}}}{{\left. {\left\| {{X_j} - {X_i}} \right.} \right\|}}\left( {1 + J\cos \left( {{\theta _{j\tau }} - {\theta _i}} \right)} \right)} \\
 - \frac{{{X_j} - {X_i}}}{{{{\left. {\left\| {{X_j} - {X_i}} \right.} \right\|}^2}}}
\end{array}
\end{equation}
\begin{equation}
\label{eq2}
{\dot \theta _i} = {w_i} + \frac{K}{N}\sum\limits_{j \ne k}^N {\frac{{\sin \left( {{\theta _{j\tau }} - {\theta _i}} \right)}}{{\left. {\left\| {{X_j} - {X_i}} \right.} \right\|}}} {\rm{.}}
\end{equation}
For i=1,...,N where N represents the population size, $ v_i $ is the spatial velocity of each swarmalator at the onset of the study, and $ \omega_i $ the natural frequency of the $i^{th}$ node. Here, $ \theta _{j\tau} $ is the contracted form of $ \theta _{j}(t-\tau) $ and $ \theta _{j} $ that of $ \theta _{j}(t) $. 

\section{Analytical tools} \label{tools}

To investigate phase synchronization of nodes in a network, Kuramoto order parameter is generally used and defined as \cite{Mirollo2012} :
\begin{equation}
\begin{aligned}
R = \left| \frac{1}{N} \sum_{j=1}^{N} e^{i \theta_j} \right|
\end{aligned}
\end{equation}
with ${\rm{R}} \in {\rm{ [0}}{\rm{,1]}}$ and $\theta_j$ the phase of the $j^{th}$ node. $R \rightarrow 1$ indicates synchronization while $R \rightarrow 0$ suggests incoherence. $R$ provides a global information on the state of the network. However, its interpretation becomes ambiguous in intermediate regimes since an intermediate value can reflect either partial synchronization or a chimeric structure.

For networks, where agents are static spatially, the characterization of chimera states can be performed using the parameters developed in 2014 by Gopal et al. in \cite{Gopal2014}. The development of that parameter is based on the definition of a local coherence constant $z_i$ for each node in the network, obtained by evaluating the variance of phase differences between a node and its immediate neighbors. Values of $z_i$  close to $0$ indicate strong local coherence, while high values indicate disordered dynamics. Based on this measurement, they proposed the Strength of incoherence index $SI$ \cite{Gopal2014}, defined by:
\begin{equation}
\begin{aligned}
SI = 1 - \frac{1}{M} \sum_{m=1}^{M} s_m,
\end{aligned}
\end{equation}
where $s_m$ is a binary function indicating whether the local group $m$ is coherent ($s_m = 1$) or incoherent ($s_m = 0$), and $M$ is the total number of sub-regions considered \cite{Gopal2014}. Thus, $SI = 0$ corresponds to synchronization, $SI=1$ to incoherence, and intermediate values could characterize chimera states. However, its application becomes limited in the case of unordered or mobile systems, such as swarmalators, for which the network topology is constantly changing \cite{mob1,mob2,mob3}.

To overcome these limitations, we use recurrence analysis tools developed in \cite{Marwan2007,Goswami2019,Lameu2018,Brandt2023,Iona2014}. This approach is based on constructing a recurrence matrix from the temporal trajectory $ \varphi $ of each oscillator of the network:

\begin{equation}
{RP_{l,m}}(\varepsilon ) = \Theta (\varepsilon  - \left\| {{\varphi _l} - {\varphi _m}} \right\|)
\end{equation}
where $ \varepsilon $ is the recurrence threshold, $l$ and $m$ represent instants in the time series $\varphi$ and $ \Theta $ is the Heaviside function. In these recurrence matrices, the black points represent recurring points, i.e. the system is revisiting an old state, and the white points are considered to be non-recurring; both axes represent time. Analysis of the structure of the recurrence matrix makes it possible to evaluate the regularity, complexity or chaotic nature of nodes. For this reason, Recurrence Quantification Analysis (RQA) measures \cite{Marwan2007,Iona2014} have been developed, including recurrence rate $(RR)$ expressed as:
\begin{equation}
RR = \frac{1}{{N(N - 1)}}\sum\limits_{i \ne j}^N {R{P_{l,m}}}.
\end{equation} 

In order to link individual dynamics to the overall state of the network, we use the Joint Recurrence Plots $(JRP)$, which are obtained by combining the outputs of
several oscillators in pairs or simultaneously \cite{Marwan2007}:

Let $ JRP_{(l,m)\pi }^{\varphi ^i,\varphi ^j} $ denote the joint recurrence matrix of trajectories $\varphi ^i$ and $\varphi ^j$, 
\begin{equation}
JRP_{(l,m)\pi }^{\varphi ^i,\varphi ^j} = \Theta\displaystyle\left(\varepsilon  - \left\| {\mathop \varphi \nolimits_l^i  - \mathop \varphi \nolimits_m^i } \right\|\displaystyle\right)\Theta\displaystyle\left(\varepsilon  - \left\| {\mathop \varphi \nolimits_l^j  - \mathop \varphi \nolimits_m^j } \right\|\displaystyle\right)
\end{equation}

A variant of the join recurrence plot is used, which is the sum of RP matrix: 
\begin{equation}
JRP_{(l,m) + }^{{\varphi ^i},...,{\varphi ^N}} = \frac{1}{N}\sum\nolimits_{k = 1}^N \Theta  ({\varepsilon _{{\varphi ^k}}} - \left\| {\varphi _l^k - \varphi _m^k} \right\|),i = 1,...N
\end{equation}

the joint recurrence rate is also given by:
\begin{equation}
JRR = \frac{1}{{{N(N-1)}}}\sum\limits_{i\ne j}^N {JR{P_{(l,m)\pi}}}
\end{equation}
We define the recurrence based synchronization index $ S_{\varphi ^i,\varphi ^j} $ as:  

\begin{equation}
{S_{\varphi ^i,\varphi ^j}} = \frac{{JR{R^{\varphi ^i,\varphi ^j}}}}{{R{R^ {\varphi ^i}}}}
\end{equation}

The visualization of pairwise synchronization measure $S$ and the histogram allows a qualitative interpretation of the overall state of the system. Areas of high density ($S \rightarrow 1$) reflect synchronization between subsets of oscillators, while lower densities ($S \rightarrow 0$) reflect uncoherent behaviors. However, for intermediate values of $S (0 < S < 1)$, the network exhibits a tendency toward coherence that strengthens as S increases. In such cases, the dynamics of the nodes become increasingly similar, though not identical.

In order to quantify this complexity, we associate the JRPs with the calculation of the Shannon entropy \cite{Shannon1948,CH36}.

\begin{equation}
    \rho  = -\sum\nolimits_a {{p_a^M}} \log ({p_a^M})
\end{equation}
where $p_a^M$ is the probability of finding an element $a$ in a matrix $M$.

$p_a^M = \frac{{\mathop n\nolimits_a }}{{\mathop {S_m}\nolimits^2 }}$  where $n_a$ denotes the number of occurrences of element $a$, and $S_m$ is the length of the square matrix $M$.
We define the measures $\rho_\pi$ and $\rho_+$ such that:
\begin{equation}
    \rho_\pi  = -\sum\nolimits_a {{p_a^{M_\pi}}} \log ({p_a^{M_\pi}})
\end{equation}
\begin{equation}
    \rho_+  = -\sum\nolimits_a {{p_a^{M_+}}} \log ({p_a^{M_+}})
\end{equation}
with $M_\pi = S$ and $M_+ = JRP_{(l,m) + }$.

Entropy measures the diversity of recurrent values and therefore the degree of complexity in the system such that high entropy value characterizes a wide variety of behaviors, while zero entropy corresponds to complete synchronize state where all recurrences are identical.

The individual analysis of a node and its contribution to the dynamics of the system is very important to improve our understanding of the network states.
We hypothesize that the chaotic activity of the network may prevent us from distinguishing oscillators that are correlated with each other. This is due to an index misalignment or a non spatial proximity. To address this, we propose evaluating the network’s coherence trend not in terms of coherence regions, but rather by identifying nodes that share the same global dynamics.
To do so, a correlation threshold based on the measures  of $S$ 
is set in order to distinguish oscillators considered to be completely independent (uncorrelated). These nodes constitute the incoherent part of the system, and their proportion is used to calculate the degree of independence $N_L$ defined as:
\begin{equation}\label{nl}
{N_L} = \frac{{N_i}}{N}
\end{equation}
where ${N}$ is the total number of nodes in the system and ${N_i}$ represents the number of nodes identified as purely independent. 

\section{Numerical results} \label{results}

For swarmalators the key parameters include spatial coupling strength $J$, phase coupling $K$,and delay $\tau$. The typical population contained between 100 and 600 agents randomly distributed in space initially in the interval $\left[-1,1\right]$, with phases randomly initialized also in $\left[-\pi,\pi\right]$. All oscillators are assumed to be identical with the same speed and natural frequency. For Colpitts oscillators, the number of neighbors $p$ is taken into account as well as the coupling coefficient $r$ for a network of 100 oscillators. The fixed parameters of each oscillator are $Vcc=15$; $\beta = 5$, $Vref=0.1$. The Runge-kutta method of order 4 (RK4) is used to integrate the differential equations governing the dynamics of the systems. 

\subsection{\textbf{Colpitts Oscillator Network}}

We used the asymptotic standard deviation $\left\langle \sigma  \right\rangle $  to select the synchronization regimes points of the oscillator network \cite{Carmago2024}. This parameter is based on the instantaneous standard deviation ${\sigma _t} $, calculated at each instant t from the dispersion of the state variables of the oscillators around their respective means.
\begin{equation}
\begin{array}{l}
{\sigma _t} = \left[ {\frac{1}{N}\sum\limits_{i = 1}^N {\left( {{{\left( {x(t) - \bar x(t)} \right)}^2} + } \right.} } \right.{\left( {y(t) - \bar y(t)} \right)^2}\\
{\left. {\left. { + {{\left( {z(t) - \bar z(t)} \right)}^2}} \right)} \right]^{1/2}},
\end{array}
\end{equation}
where $\bar{x}(t)$, $\bar{y}(t)$ and $\bar{z}(t)$ represent the averages of the variables $x_i$, $y_i$ and $z_i$ across the entire network at time $t$.
Over an observation period T, the temporal mean of this standard deviation, provides information on the overall coherence of the system.
\begin{equation}
\langle \sigma \rangle = \frac{1}{T}\sum_{t=t_0+1}^{t_0+T} \sigma_t.
\end{equation}

A value of $\left\langle \sigma  \right\rangle  = 0$ indicates complete synchronization of all oscillators, while higher values indicate partially synchronized or disordered regimes.

The results obtained are presented in the form of a two-dimensional diagram $\left(r,p\right)$, representing respectively the coupling coefficient and the number of neighbors. This diagram allows us to visualize the areas of synchronization and desynchronization, as well as the transitions between the different collective states of the system.
\begin{figure}[htbp]
    \centering
    \includegraphics[width=8.5cm,height=6cm]{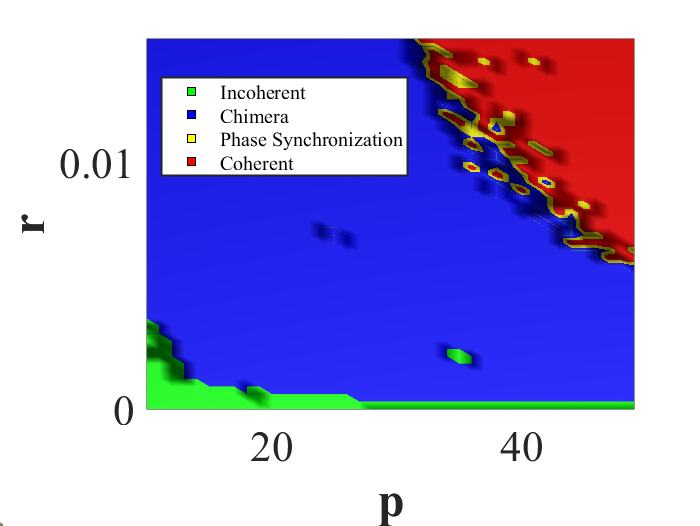}
    \caption{Map of synchronization regimes for a Colpitts network of $N=100$ nodes as a function of the number of nearest neighbors $p$ and the coupling coefficient $(0 \le r \le 0.015)$. The green area denotes disorder (desynchronization), the blue area corresponds to chimera states, the red area to complete synchronization, and the yellow-shaded regions indicate phase-synchronization. }
   \label{fig:placeholder}
\end{figure}

\subsection*{Synchronization and Desynchronization}

For very low values of the coupling coefficient there is almost no dependence between the individual dynamics of the oscillators. Then each recurrence matrix has its own pattern and structure, reflecting an autonomous and uncorrelated evolution. This is the case in Fig.\ref{fig:rp42}, where four randomly chosen nodes $57$, $66$, $86$ and $99$ show completely different structures of their recurrence matrices. However, while increasing the values of the coupling coefficient $r$, the system variables evolve coherently and analysis reveals similar structure in the recurrence plots of all the oscillators, reflecting synchronized dynamics on the scale of the network (See Fig.\ref{fig:rp97}, where nodes $76$ and $97$ show almost identical recurrence matrices). 

\begin{figure}[htbp]
\centering
\includegraphics[width=8.5cm,height=8cm]{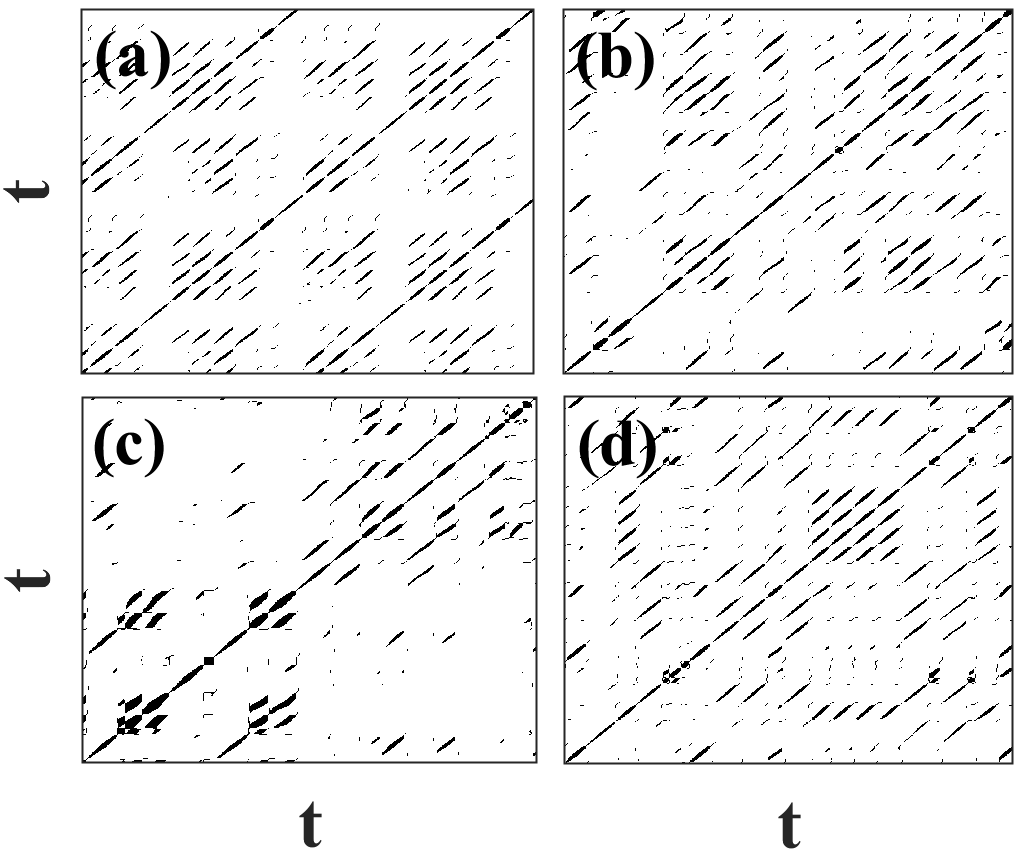}	
\caption{Recurrence plots of four randomly chosen nodes $57$, $66$, $86$ and $99$ taken from incoherence in the network showing different structures. The recurrence threshold is chosen so that the minimum recurrence rate is close to $\textbf{$RR \simeq 5\% $}$ with $p=15$ and $r=0$.}
\label{fig:rp42}
\end{figure}

\begin{figure}[htbp]
\centering
\includegraphics[width=8.5cm,height=4.5cm]{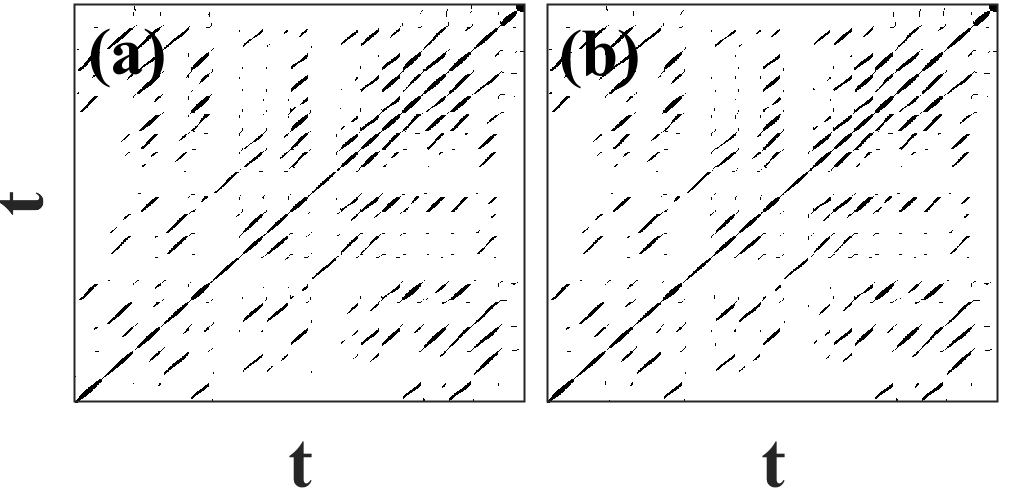}	
\caption{Recurrence plots of two randomly chosen nodes $76$ and $97$ taken from synchronization state presenting almost identical structures. $p=40$, $r=0.015$ and $RR \simeq 5\% $.}
\label{fig:rp97}
\end{figure}

Whether in totally disordered or synchronized regime, the recurrence matrices reveal irregular patterns characteristic of chaotic dynamics.
Even in partially coherent regimes, this instability persists, indicating that global synchronization does not suppress chaos, but rather reflects temporal coordination between oscillators \cite{Kiss2005}.

From the recurrence synchronization measure matrix, it is possible to determine the overall state of the system by comparing the joint recurrence rates between the different pairs. This approach allows to assess the degree of similarity between trajectories and to detect any collective structures. The matrices, presented in Fig.\ref{fig:jrpdensity}, correspond to different collective states identified previously, illustrating the incoherent Fig.\ref{fig:jrpdensity} (a) and (b) and synchronization Fig.\ref{fig:jrpdensity} (c) and (d) regimes observed within the network.
\begin{figure}[htbp]
\begin{tabular}{cc}
(a)&(b)\\
\includegraphics[width=4.4cm,height=4cm]{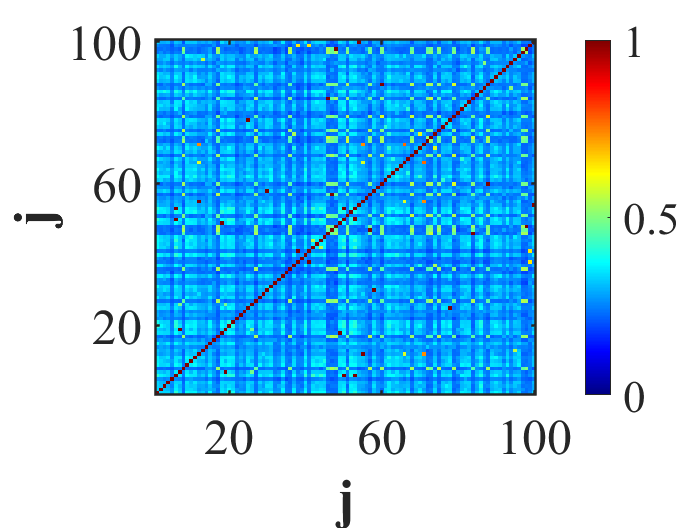}&
\includegraphics[width=3.95cm,height=4cm]{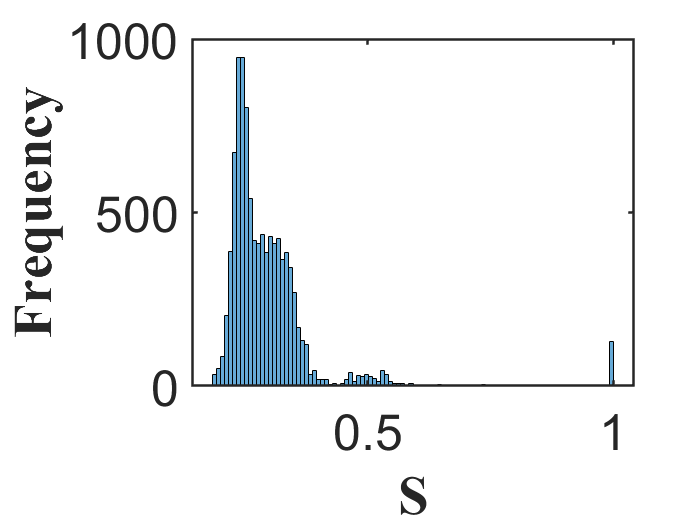} \\
(c)&(d)\\
\includegraphics[width=4.4cm,height=4cm]{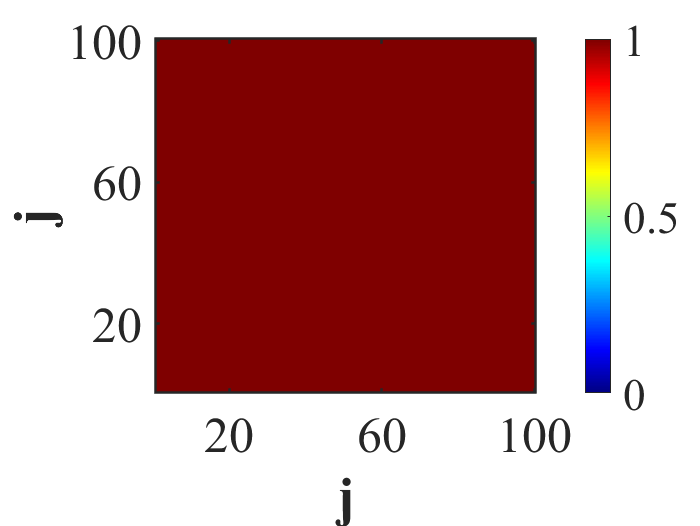}&
\includegraphics[width=3.95cm,height=4cm]{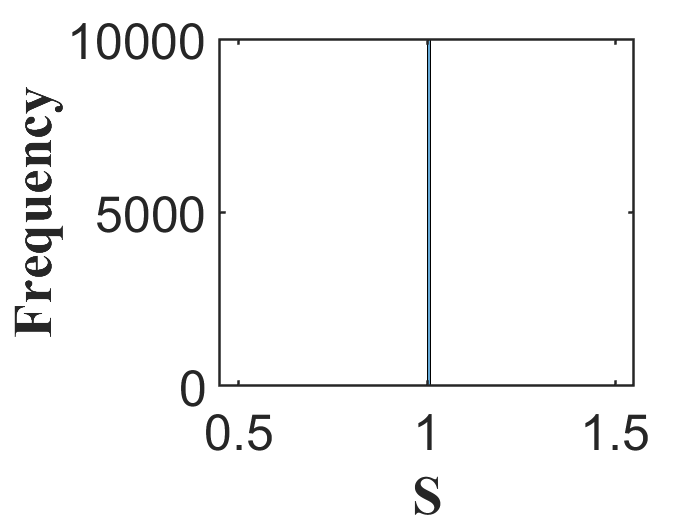}

\end{tabular}
\caption{Recurrence synchronization measure of Networks . Incoherent state for $r=0$:
Fig.\ref{fig:jrpdensity} (a) shows a wide variety of weak values of Recurrence synchronization measure and Fig.\ref{fig:jrpdensity} (b) presents the distribution of the values mostly located in the weak values of $S$. Synchronization for $r=0.015$ is obviously presented by the unique value of S Fig.\ref{fig:jrpdensity} (c) corresponding to unique peak at $S=1$ in the histogram Fig.\ref{fig:jrpdensity} (d). }
\label{fig:jrpdensity}
\end{figure}

Analysis of the Recurrence synchronisation measure matrix reveals contrasting dynamic signatures depending on the network regime. In disordered mode, the Recurrence synchronisation measure matrix highlights generally widely distributed weak correlations between oscillators (See Fig.\ref{fig:jrpdensity} (a)). The corresponding histogram (Fig.\ref{fig:jrpdensity} (b)) shows the distribution in many peaks mostly located in the weak values. These observations are characteristic of a weakly coupled chaotic system, in which the intrinsic dynamics of each oscillator dominate interactions, preventing the emergence of collective coordination.
Conversely, in the fully synchronized regime, the JRP density matrix appears uniformly saturated, indicating maximum correlation between all pairs of nodes. The associated histogram shows a single peak centered on the maximum value $S=1$, reflecting total network coherence. This behavior corresponds to complete collective locking of the oscillators, induced by the dominant effect of coupling.

\subsection*{Chimera State}

For intermediate values of coupling, we observe a dynamic separation within the network: some oscillators show marked coherence, while others retain an incoherent behavior. This stable coexistence of coherent and incoherent sub-populations, despite the structural homogeneity of the network, is the characteristic signature of a chimera state. Analysis tools based on joint recurrence reveal this hybrid structure.
 \begin{figure*}[htbp]
 \centering
 \begin{tabular}{ccc}
 (a)&(b)&(c)\\
 \includegraphics[width=6cm,height=6cm]{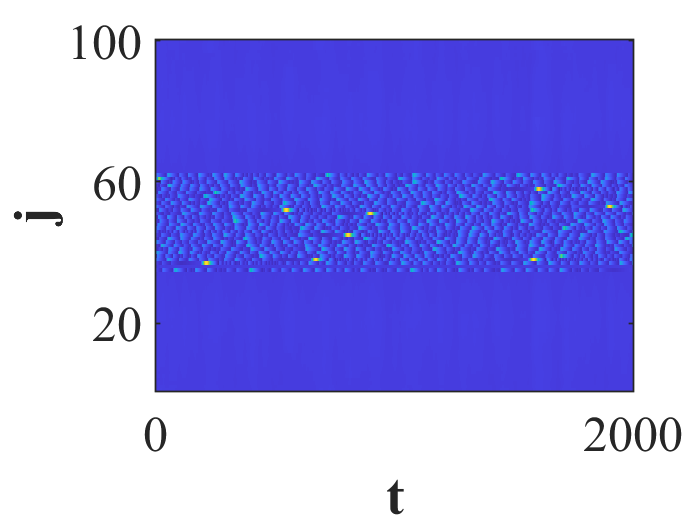}&
 \includegraphics[width=7cm,height=6cm]{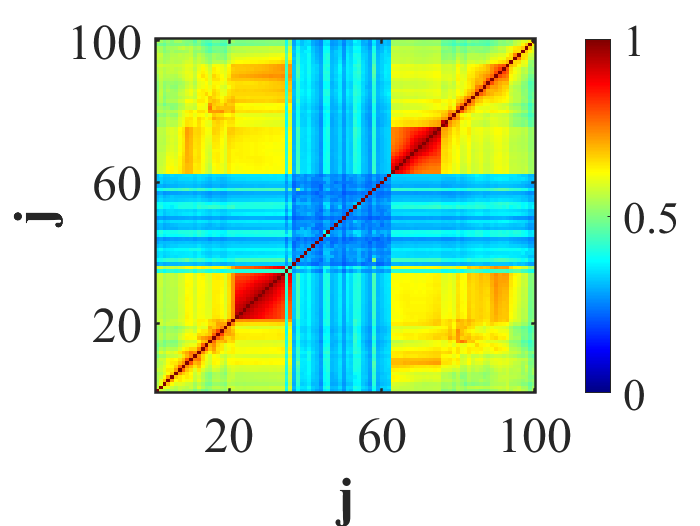}&
 \includegraphics[width=5cm,height=6cm]{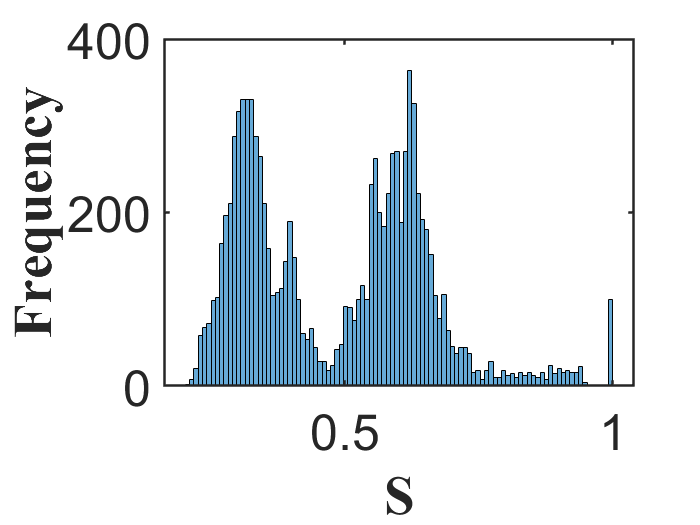}\\
($a_1$)&($a_2$)&($a_3$)\\
 \includegraphics[width=0.325\textwidth]{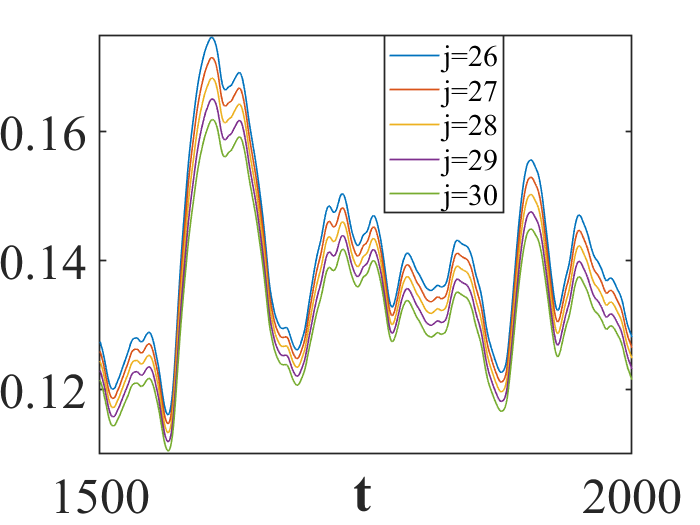}&
 \includegraphics[width=0.325\textwidth]{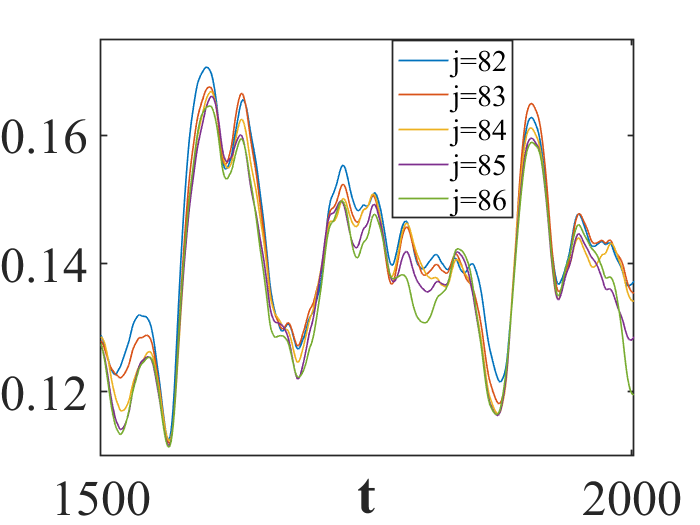}&
 \includegraphics[width=0.325\textwidth]{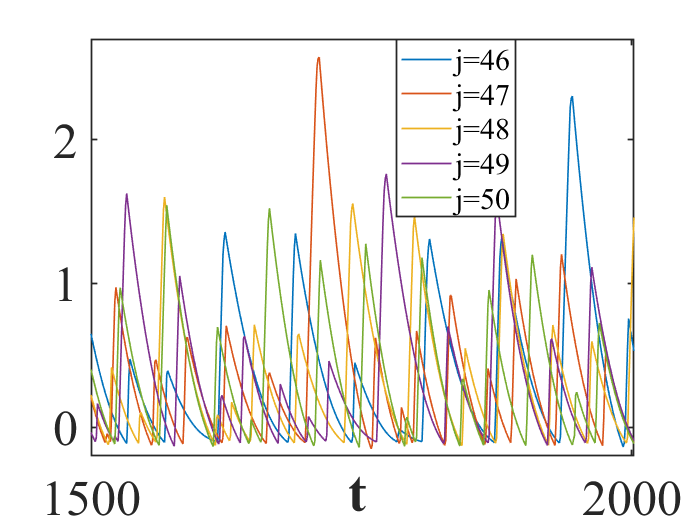}
 
 \end{tabular}
 \caption{Chimera state ($p=40$, $r=0.001$ and $RR = 1\%$ ). Fig.\ref{fig:jrpdensitychim}.(a) shows the spatiotemporal diagram of the network and exhibits two coherent regions (where some nodes are selected to show their relative behaviors in Figs.\ref{fig:jrpdensitychim}.($a_1$) and ($a_2$)) separated by an incoherent one (Fig.\ref{fig:jrpdensitychim}.($a_3$)). While the recurrence synchronization measure matrix $S$ shown in Fig.\ref{fig:jrpdensitychim}.(b) helps to evaluate the closeness between pairs according to the color bar, its corresponding histogram in Fig.\ref{fig:jrpdensitychim}.(c) evaluates the density of each $S$'values. It comes out two intervals of high frequency. The large majority of nodes are: incoherent for around $0.2\leq S\leq 0.4$ and are intermediately coherent for around $0.55\leq S\leq 0.7$. Few pairs of nodes present higher values of $S > 0.7$ but less than $1$. $S=1$ corresponds to the diagonal elements of the matrix, i.e., the self-similarity of each oscillator with itself. Nevertheless, despite the presence of these different coherence levels, no complete synchronization is observed among the oscillators in the network.}
 \label{fig:jrpdensitychim}
 \end{figure*}

 In this intermediate regime, the Recurrence synchronization measure matrices reveals a clearly heterogeneous structure. Certain areas corresponding to subgroups of oscillators exhibiting mutual correlation coexist with more diffuse regions, where coherence is significantly weaker.

  Fig.\ref{fig:jrpdensitychim} shows that, in the coherent regions, the oscillators are not perfectly synchronized; however, they exhibit very similar amplitudes and phases (see Figs.\ref{fig:jrpdensitychim}.($a_1$) and ($a_2$)). In contrast, the incoherent regions are characterized by significant oscillation amplitudes and phases dispersion, indicating a loss of collective coherence (see Fig.\ref{fig:jrpdensitychim}.($a_3$)).
 The associated histogram highlights two distinct correlation peaks, illustrating the simultaneous presence of two major groups based on their density: the first group in incoherence domain for $ S\in \left[0.2,0.4\right]$ and intermediately coherence area for $ S\in \left[0.55,0.7\right]$. These two intervals contain the majority of the nodes constituting our network. Having simultaneously high density of low ($<0.5$) and high values ($>0.5$) of $S$ reflects the spontaneous formation of correlated agents within a globally chaotic network, a phenomenon characteristic of chimera states. From a physical point of view, this regime results from a delicate balance between the intrinsic tendency towards chaos of isolated oscillators and the collective attraction effect generated by coupling, which promotes coherence in certain regions of the network.\\

It should be noted that the number of closest neighbours taken into account in the coupling has a decisive influence on the width of the system's synchronization band. Indeed, as the number of neighbours increases, the synchronization range widens significantly, reflecting better propagation of dynamic information across the network. This behaviour can be explained by the fact that more extensive coupling promotes faster diffusion of local disturbances and more effective harmonization of the phases and amplitudes of the oscillators, thereby strengthening the collective coherence of the system.\\

In order to better understand the internal structure of the collective states observed, we introduced the degree of independence $N_L$, which measures the proportion of dynamically independent oscillator groups within the network. This parameter is calculated from the number $N_i$ of groups of nodes identified as independent, determined according to a correlation threshold defined from the distribution of the Recurrence synchronization measure $S$.

The independence threshold $\epsilon$ is chosen adaptively, around the average of the overall correlations of the system $\mu$, taking into account the associated standard deviation $\sigma_c$. This approach makes it possible to adjust the sensitivity of the criterion according to the dynamic regime under consideration, whether it is a synchronized, disordered or chimerical state. $\epsilon=\mu \pm \alpha\sigma_c $.

The size of the independent groups thus identified also provides essential information: it indicates the level of fragmentation of the system. Small group sizes reflect marked independence and high disorder, while larger group indicates the presence of a persistent coherent structure within the overall dynamics (See Fig.\ref{grp_number}).

This treatment therefore makes it possible to link the overall value of the degree of independence to the actual fragmentation of the network and to quantitatively interpret the coexistence of coherent and incoherent areas, which are characteristic of chimeric states.

\begin{figure*}[htbp]
\centering
\begin{tabular}{ccc}
(a).$N_L=0.65$ & (b).$N_L=0.3$  & (c).$N_L=0.01$ \\
 \includegraphics[width=0.325\textwidth]{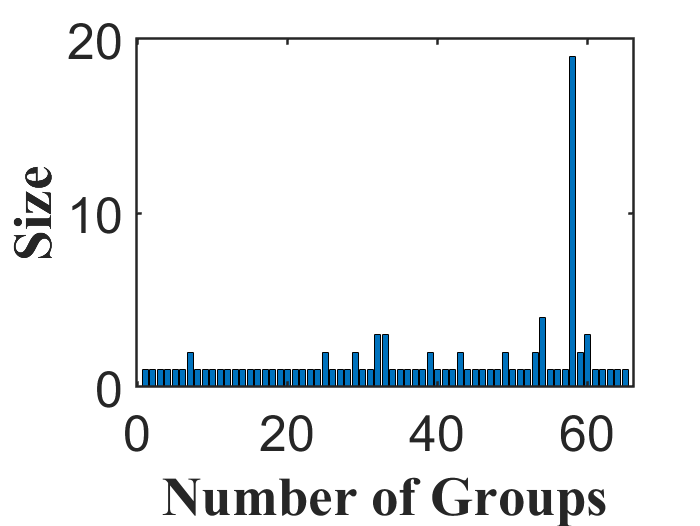}&
 \includegraphics[width=0.325\textwidth]{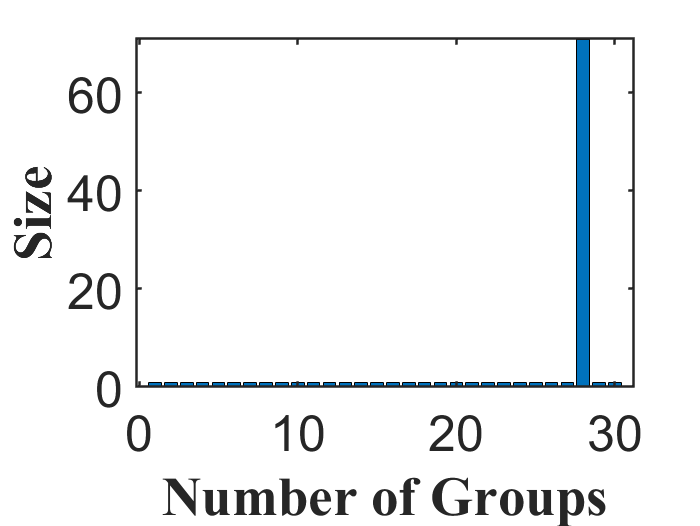}&
 \includegraphics[width=0.325\textwidth]{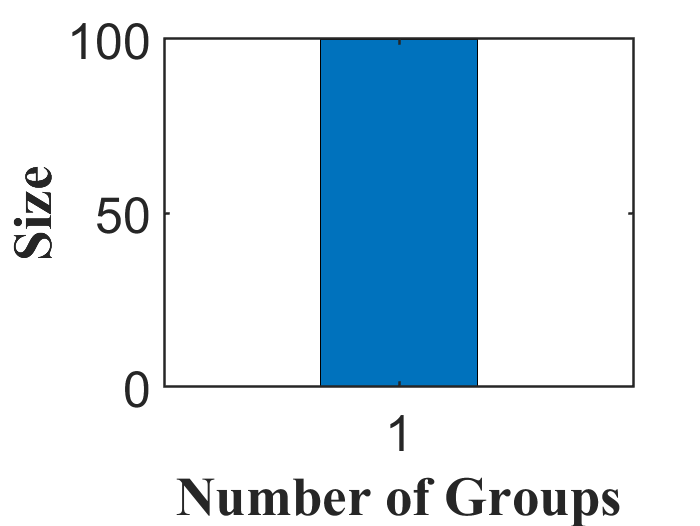}\\
 $(r;p)=(0;15)$ & $(r;p)=(0.001;40)$  & $(r;p)=(0.015;40)$ 
 \end{tabular}
\caption{Distribution of independent groups and their sizes for different collective states in a network of 100 coupled Colpitts oscillators: (a) disordered state $(N_L=0.65, r=0, p=15)$; (b) chimera state $(N_L=0.3, r=0.001, p=40)$; and (c) synchronized state $(N_L=0.01, r=0.015, p=40)$. The disordered state is characterized by many small groups, indicating weak dynamical correlations. The chimera state exhibits the coexistence of one dominant large group and several smaller groups, reflecting the simultaneous presence of coherent and incoherent oscillator populations. In the synchronized state, all oscillators belong to a single group of size $N=100$, corresponding to complete synchronization.}
\label{grp_number}
\end{figure*}

Here in Fig.\ref{grp_number}, the number of groups provides information about the number of clusters ( a cluster is an ensemble of correlated nodes) and the size is the number of elements in each cluster.

The distributions observed broadly reflect the collective regimes identified previously. In the disordered state Fig.\ref{grp_number}.(a), the distribution is highly fragmented and contains many small groups together with a larger dominant group, indicating the presence of local dynamical correlations without global organization. In the chimera state Fig.\ref{grp_number}.(b), a large coherent group coexists with several small groups, revealing a clear separation between coherent and incoherent oscillator populations. Finally, in the synchronized state Fig.\ref{grp_number}.(c), all oscillators belong to a single group, corresponding to complete synchronization. These results show that the group-size distribution provides a simple way to distinguish disordered, chimera, and synchronized regimes.\\

For the fixed value of the number of neighbours $p=40$, we studied the evolution of the network's dynamic regimes as a function of the coupling coefficient $r$. This analysis is based on three complementary quantities: the entropies $ \rho $ of the recurrence matrices, the degree of independence $N_L$, and the incoherence indix $SI$ proposed by Gopal et al \cite{Gopal2014}. These parameters make it possible to simultaneously characterise the overall complexity, the proportion of autonomous elements, and the local coherence within the network.
\begin{figure*}[htp]
\includegraphics[width=17.5cm,height=10cm]{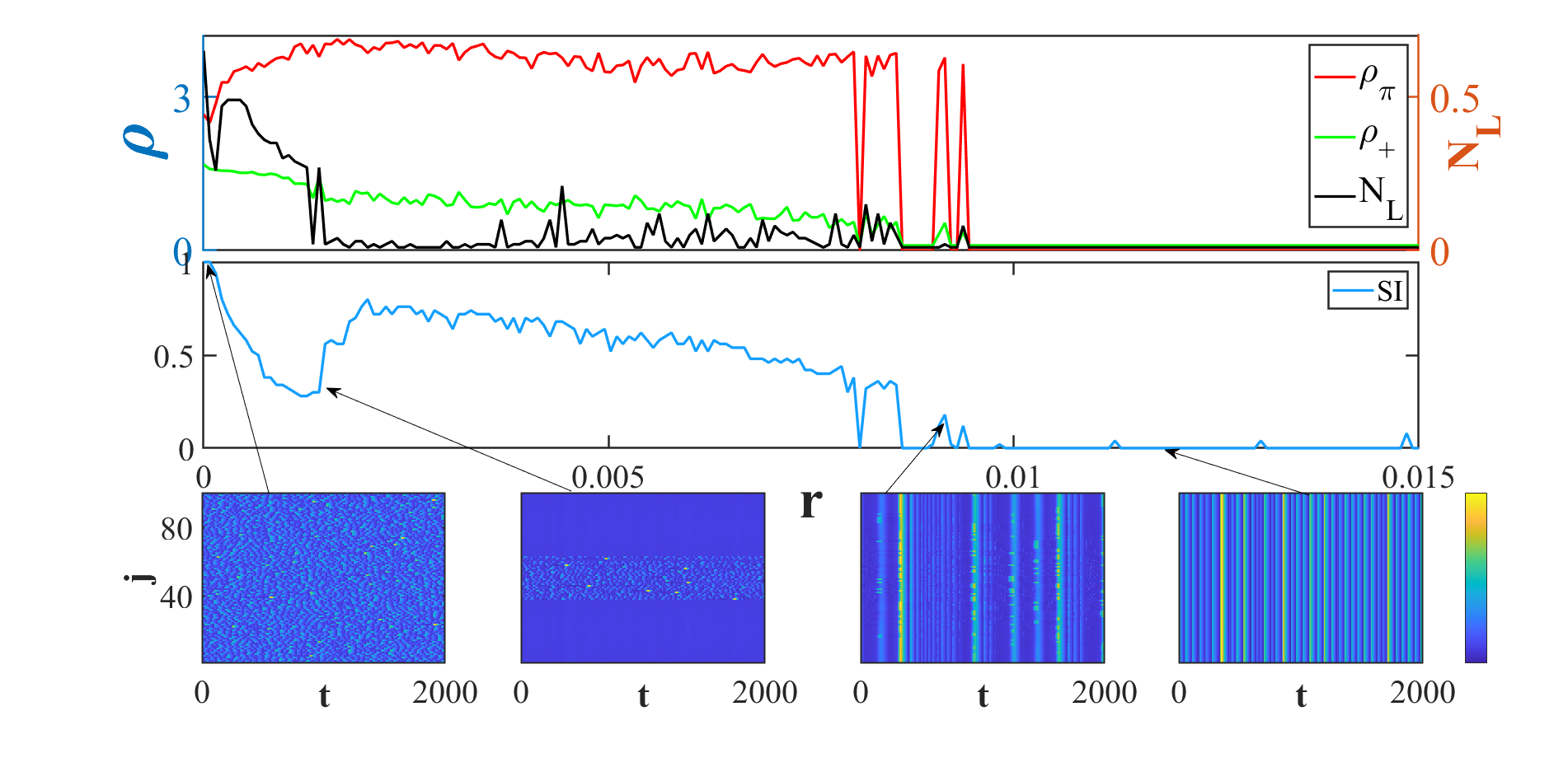}
\caption{Evolution of the recurrence-based measures $(\rho_\pi, \rho_+)$, and the independence degree $(N_L)$, together with the Strength of Incoherence $(SI)$, during the transition from a disordered state to complete synchronization through chimera states ((p=40)). The bottom panels show the corresponding spatiotemporal patterns for selected values of the coupling parameter $(r)$. As $(r)$ increases, $(N_L)$ decreases from values close to unity toward zero, indicating the progressive reduction in the number of dynamically independent oscillator groups. Simultaneously, the entropy measures $(\rho_\pi)$ and $(\rho_+)$ capture changes in the distribution of dynamical correlations, while $(SI)$ exhibits a similar transition from incoherence to synchronization. The consistency between $(N_L)$, $(\rho_\pi)$, $(\rho_+)$, and $(SI)$ demonstrates that recurrence-based measures provide a reliable characterization of disordered, chimera, and synchronized states.}
\label{fig:entro_var_colpitt}
\end{figure*}

When coupling is very weak, the degree of independence $N_L$ reaches its highest values, indicating that the majority of oscillators evolve quasi-autonomously, without any notable correlation with the rest of the network. Gopal's indice corroborate this observation: the strength of incoherence $SI$ then reaches its maximum value, reflecting global disorder and the absence of stable collective structure

As the coupling coefficient $r$ increases (see Fig.\ref{fig:entro_var_colpitt}), the strength of incoherence gradually decreases towards zero, corresponding to the values of r where the system reaches a state of complete synchronization. At the same time, the degree of independence $N_L$ decreases significantly, but shows notable fluctuations in the intermediate zone. These variations reflect the system's transition through several partially coherent dynamic regimes, characteristic of chimera states, before stabilising at a minimum value when global synchronization is achieved.

The evolution of entropies, $\rho_\pi$ and $\rho_+$, in Fig.\ref{fig:entro_var_colpitt} shows a gradual transition from total disorder to global synchronization, passing through chimera states. High and unstable entropy values characterize disordered and chimeric regimes, while a sharp fall towards zero indicates complete synchronization. This reveals that entropy $\rho$ of a recurrence matrix is a sensitive and relevant indicator for detecting and characterizing state transitions in complex coupled systems.

After examining in detail the Colpitts oscillator network, whose fixed structure allows for clear indexical organization of nodes, we were able to identify and characterize the main dynamic regimes observed: synchronization, desynchronization and intermediate chimera-like states. The joint analysis of the incoherence strength proposed by Gopal et al., and methods derived from recurrence analysis, revealed with precision the respective signatures of coherence and decoherence in the network.

The introduction of the $N_L$ independence degree then provided a complementary and robust measure, making it possible to quantify the proportion of completely autonomous oscillators in each regime. The consistency between the diagnoses obtained by these different indicators confirms the relevance and validity of the methodological approach adopted for the identification of chimera states.

Based on these results, it now makes sense to extend this analysis to the swarmalator system, whose dynamics differ significantly from those of the Colpitts network. Unlike the latter, swarmalators have an evolving topology, with the positions of agents continuously changing in space over time. This mobility makes their indexical structure intrinsically disordered, posing an additional challenge for the detection and characterisation of coherent and incoherent states. Applying the tools developed previously in this new framework will enable us to evaluate their ability to identify chimeras in mobile and topologically dynamic systems.

\subsection{Delayed Swarmalator }
The study of the swarmalator system with delay reveals a diversity of collective behaviours linked to the interaction between spatial and phase dynamics. Two types of synchronous states have been identified: Static Sync, corresponding to complete phase synchronisation, and Static Ring Sync, in which the swarmalators organise themselves into concentric rings with local synchronisation.
At the same time, active regimes have been observed, notably Active Async, characterised by total disorder, and Boiling State, marked by the coexistence of a coherent central region and a disordered periphery. A final regime, known as Boiling Chimera State, illustrates a typical hybrid state where coherence and decoherence coexist within the same system.\\ 

\subsection*{Sync States}


Two synchronized static states were observed in the swarmalator system with delay. The first, called Static Sync  Fig.\ref{scatter_100s}(e), corresponds to a state of global synchronization in which all swarmalators share the same phase. This state, already well documented in the context of swarmalators without delay, reappears here for positive phase coupling values, confirming the robustness of this dynamic regime when delay is introduced into interactions.

The second state, Static Ring Sync  Fig.\ref{scatter_100s}(a), emerges from certain negative phase coupling values. In this regime, the swarmalators organize themselves spatially on a concentric circle while maintaining local phase synchronization. Agents belonging to the same ring maintain the same phase, while a small phase difference (modulo $2\pi$) remains between adjacent  \cite{lambu}. These two regimes, although both synchronous, differ in their spatial structure and the nature of their coherence.

This subtle difference could not be highlighted by the order parameter, whose values remain high and very similar in both cases, reflecting a high degree of overall synchrony. However, the Recurrence synchronization measure $(S)$ matrix reveals this distinction Fig.\ref{scatter_100s}: for Static Sync, $S$ is uniformly dense Fig.\ref{scatter_100s}(f), indicating perfect correlation between all nodes; while for Static Ring Sync, it shows slight variations in density Fig.\ref{scatter_100s}(b) , indicating quasi-synchronization, with very high but not identical correlations between the different groups of agents, which favors a non-zero variable entropy for the recurrence matrices of this state.

Finally, the evaluation of the degree of independence $N_L$ in these regimes confirms the absence of completely autonomous nodes. In both cases, we have a minimum degree (one group  Fig.\ref{scatter_100s}(d) and  Fig.\ref{scatter_100s} (h)), which indicates that the entire system forms a single coherent group, characteristic of a globally synchronized state.

\begin{figure*}[htbp]
\centering
\begin{tabular}{cccc}
(a)&(b)&(c)&(d)\\
\includegraphics[width=4cm,height=4cm]{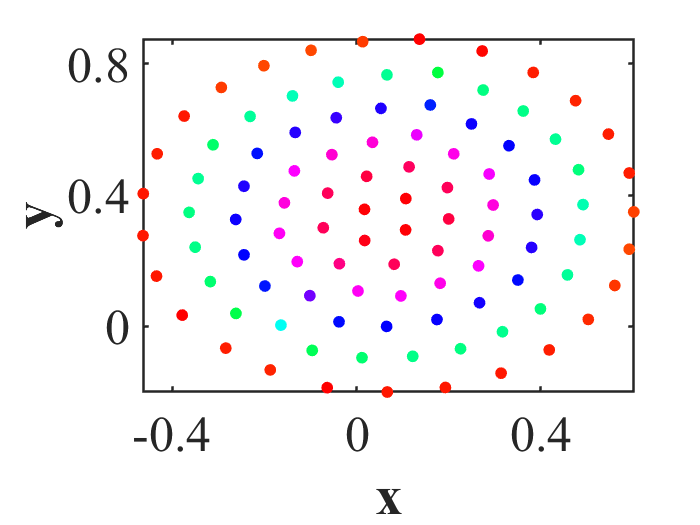}&
\includegraphics[width=4.5cm,height=4cm]{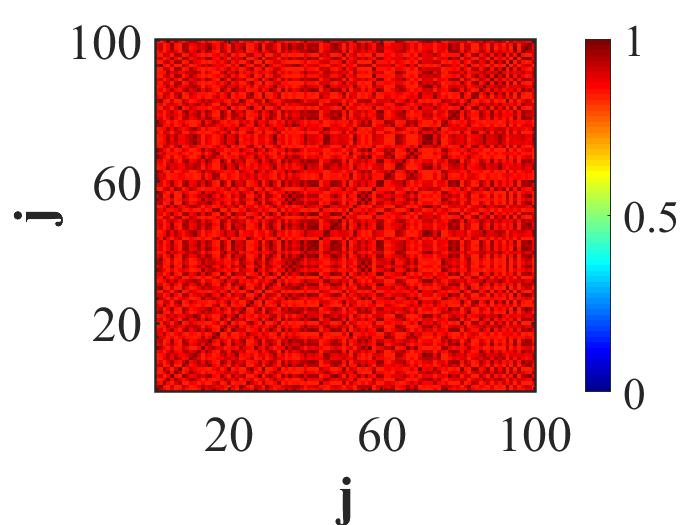}&
\includegraphics[width=4cm,height=4cm]{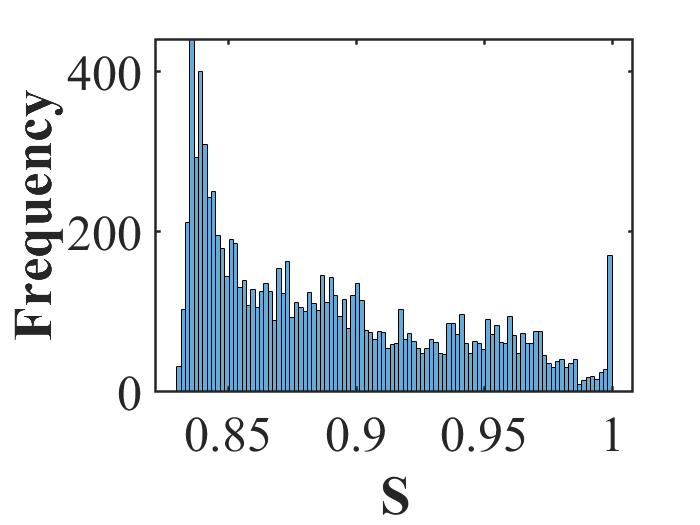}&
\includegraphics[width=4cm,height=4cm]{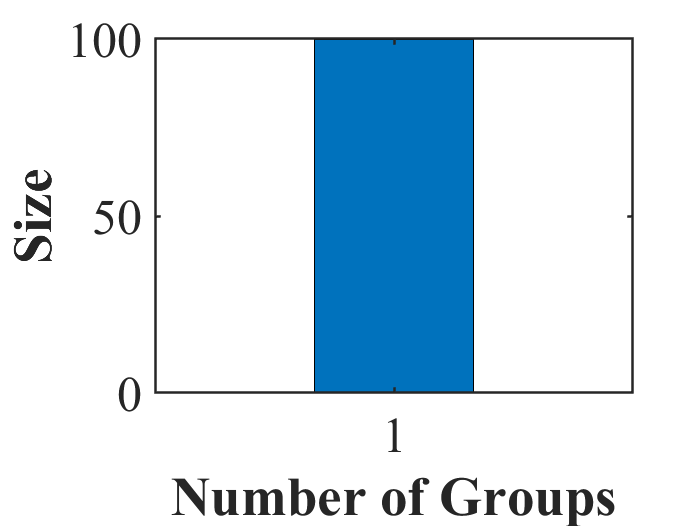}\\
(e)&(f)&(g)&(h)\\
\includegraphics[width=4cm,height=4cm]{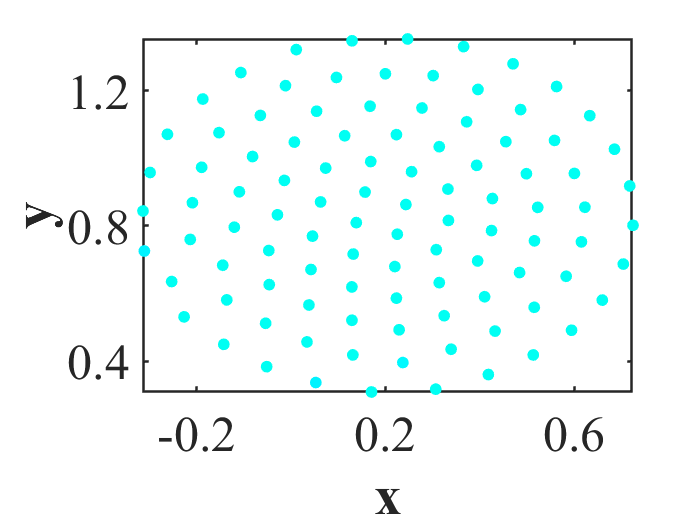}&
\includegraphics[width=4.5cm,height=4cm]{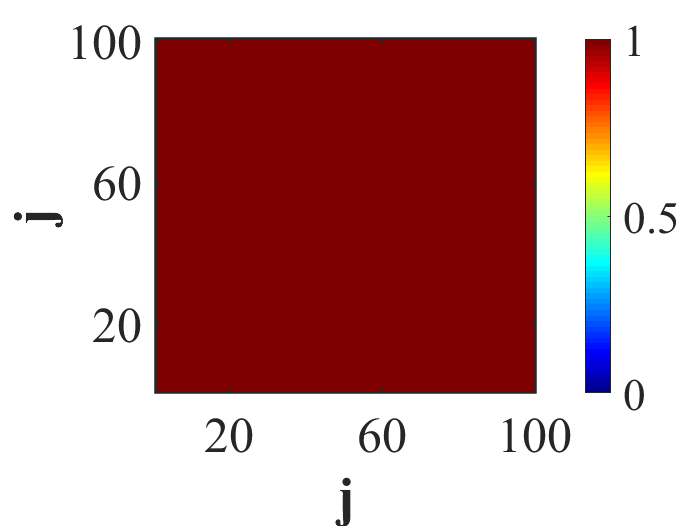}&
\includegraphics[width=4cm,height=4cm]{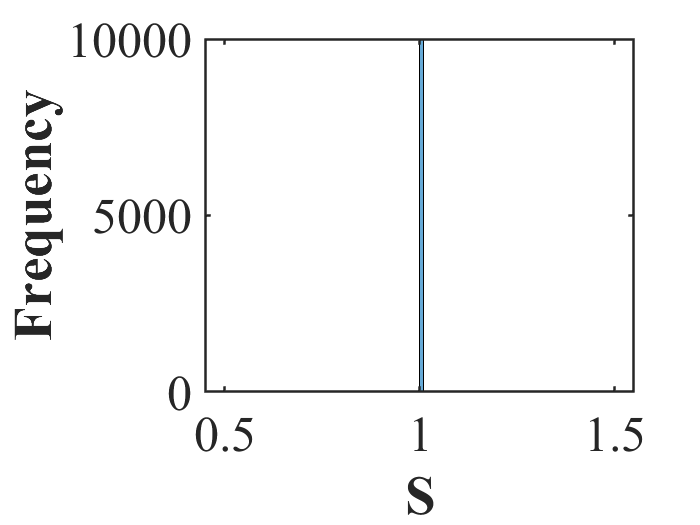}&
\includegraphics[width=4cm,height=4cm]{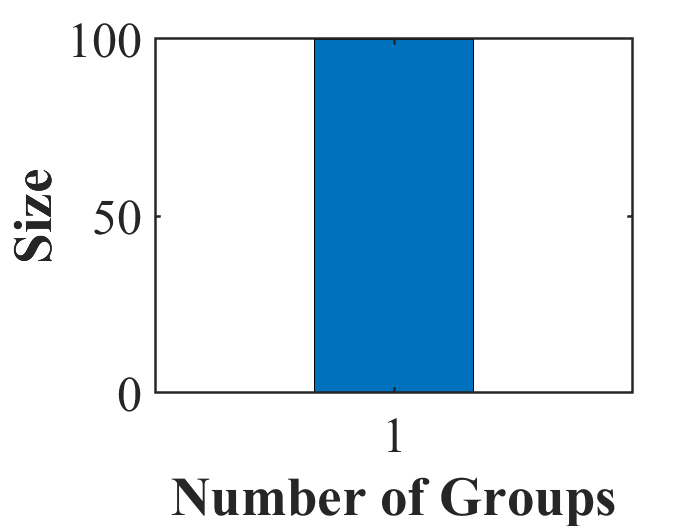}
\end{tabular}	
\caption{Long-term collective behaviour with delay.(a)"Static ring sync state" for ($J,K,\tau$)=(-1,-0.95,4.5); (e)"Static sync state" for($J,K,\tau$)=(1,0.2,3); (b) and (f) are recurrence synchronization measure matrix of these state and (c) and (g) their histogram respectively. (d), (h) number of independent group for these coherent networks. }
\label{scatter_100s}
\end{figure*}

\subsection*{Active Async:}

In a short phase coupling interval, we can see the swarmalator network evolves towards a disordered active regime, that we can call Active Async.
In this state Fig.\ref{scatter_100ds}.(a), the agents exhibit highly dynamic and uncorrelated trajectories: their positions are continuously dispersed in space, while their phases oscillate without any stable relationship to those of the other nodes. No persistent spatial or temporal structure is observed, and the phase distributions remain random.

The S matrix associated with this regime are characterized by low and irregularly distributed density values Fig.\ref{scatter_100ds}.(b),, reflecting the lack of synchronization between swarmalators. The density histogram shows a single peak concentrated around low values, confirming the very low probability of coincidence between recurrence events at different nodes.
This absence of correlated structure is typical of fully developed chaotic dynamics, where each agent follows its own evolution without collective constraints.

The calculation of the degree of independence confirms this observation: it reaches values close to 1, indicating that all swarmalators behave autonomously. This measurement thus highlights a state of global desynchronization, in which local coherence is non-existent and interactions between agents are no longer sufficient to stabilize a collective organization. This absence of large correlated groups is a hallmark of a highly incoherent state.

\begin{figure*}[htbp]
\centering
\begin{tabular}{cccc}
(a)&(b)&(c)&(d)\\
\includegraphics[width=4cm,height=4cm]{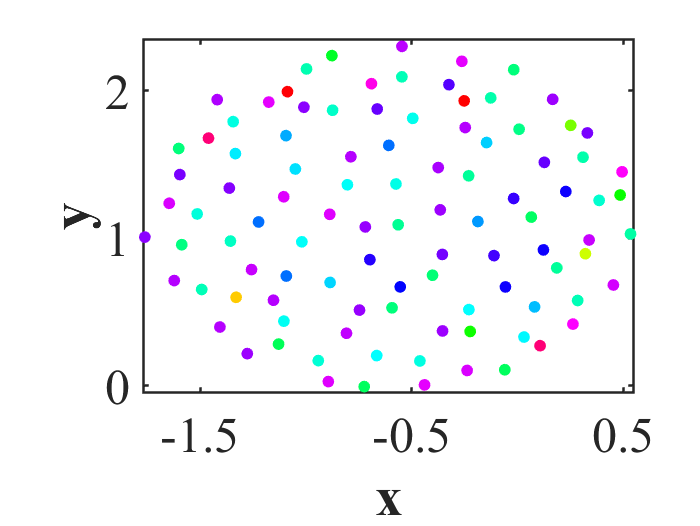}&
\includegraphics[width=4.5cm,height=4cm]{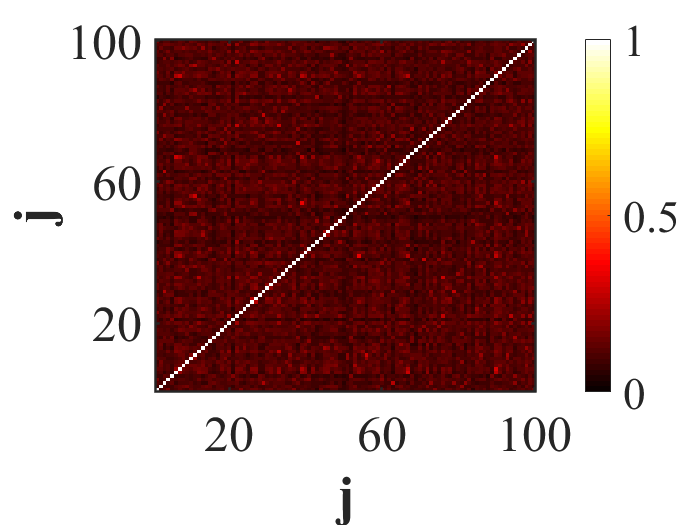}&
\includegraphics[width=4cm,height=4cm]{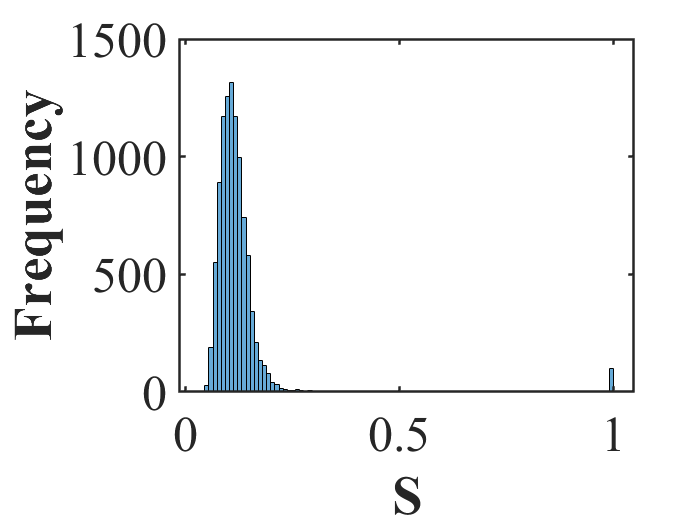}&
\includegraphics[width=4cm,height=4cm]{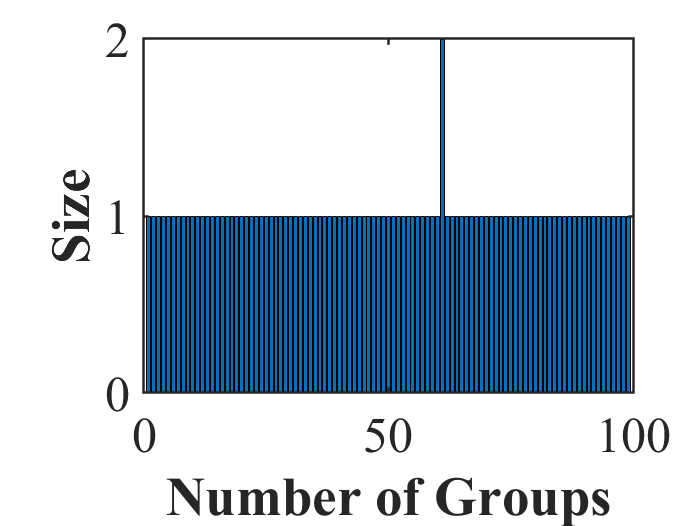}
\end{tabular}	
\caption{Active async state.(a)"scatter" shows a disordered phase distribution across the entire network, (b) Recurrence synchronization matrix characterized by a low density of synchronization values, mainly distributed around 0.2, as shown by the histogram in (c). (d) Group-size distribution displaying nearly 100 independent groups $(N_L=0.99)$, indicating that almost every oscillator forms an isolated dynamical unit.}
\label{scatter_100ds}
\end{figure*}

\subsection{Boiling State}

Between fully synchronous regimes Fig.\ref{scatter_100s} and complete active disorder Fig.\ref{scatter_100ds}, the swarmalator system with delay presents a particularly interesting intermediate state, called the boiling state Fig.\ref{Boiling}.
In this regime, some of the swarmalators, generally those located in the center of the group, remain organized and coherent, while the periphery remains in constant turmoil, with no stable correlation between phases or positions. This coexistence of coherent and incoherent zones is the hallmark of a chimera state. There are two variants: the boiling state Fig.\ref{Boiling}.(a).,
 
(Observed for coupling values $K$ between -1.3 and -0.5 for fixed values of parameter (J,$\tau$)=(1,3).), and the boiling chimera state Fig.\ref{Boiling}.(b) where the internal phase synchronized state change periodically \cite{lambu}.

	\begin{figure}[htbp]
		 	\centering
		 		\begin{tabular}{cc}
		 		(a)&(b)\\
		 	{\includegraphics[width=4.15cm,height=4.15cm]{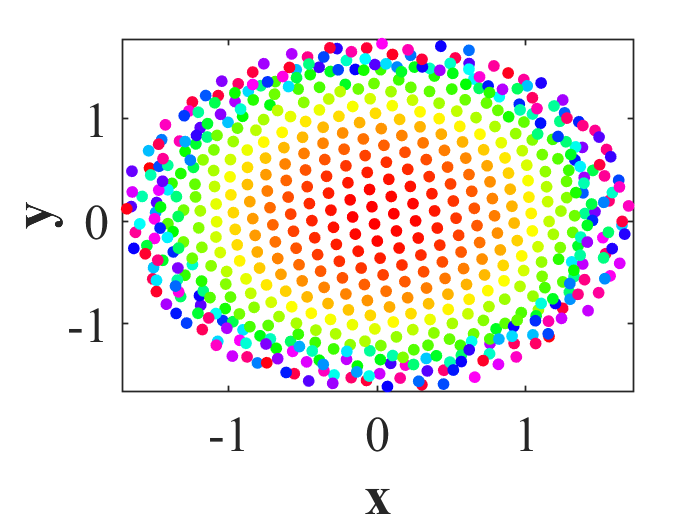}}&
		 	{\includegraphics[width=4.15cm,height=4.15cm]{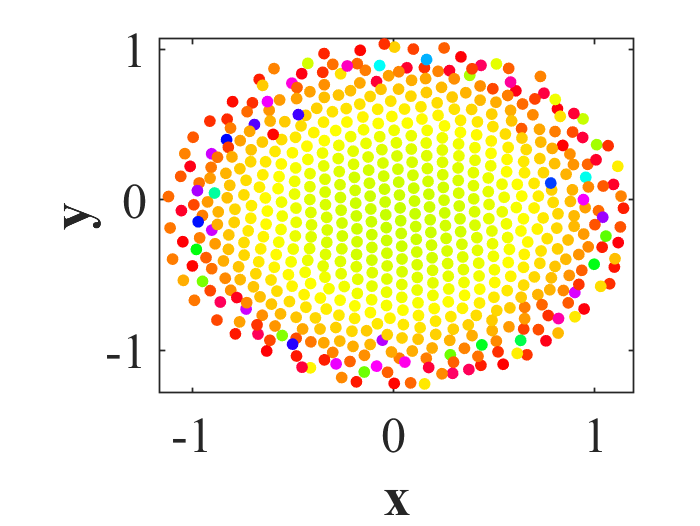}}
		 	\end{tabular}
		 \caption{Boiling state (a).(J,K,$\tau $)=(1,-0.85,3) and boiling Chimera state (b).(J,K,$\tau $)=(0.1,-1,1.95). $N= 600$. } 
		 	\label{Boiling}
		 	
	\end{figure}
The Recurrence synchronization measure matrices ($S$) associated with this regime clearly reveal this duality, which is even more pronounced in the boiling chimera state. The density histogram shows a bimodal distribution, indicating the simultaneous presence of two dynamic groups: one correlated, the other independent.

Analysis of the degree of independence also highlights this hybrid organization. The values obtained are intermediate between the extreme states (sync and async) presented above and confirmed by the analysis of the size of independent groups, indicating that part of the system remains coherent (a large group), while the other part retains an autonomous dynamic (independent nodes). 		

\begin{figure*}[t]
	 \centering
	 \begin{tabular}{cccc}
     (a)&(b)&(c)&(d)\\
\includegraphics[width=4cm,height=4cm]{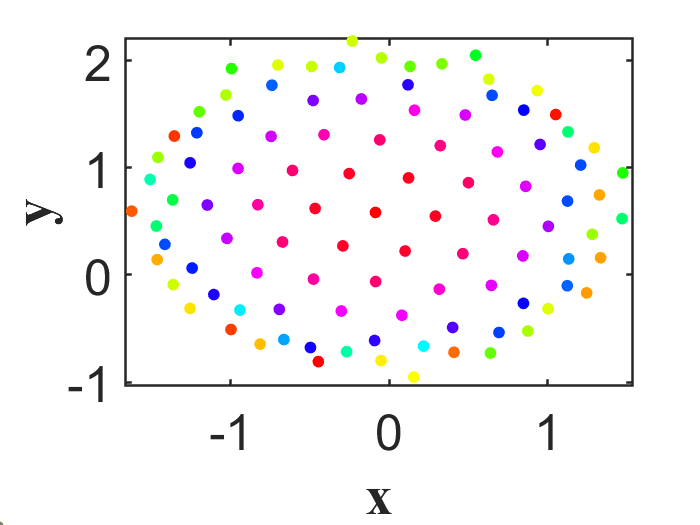}&
\includegraphics[width=4.5cm,height=4cm]{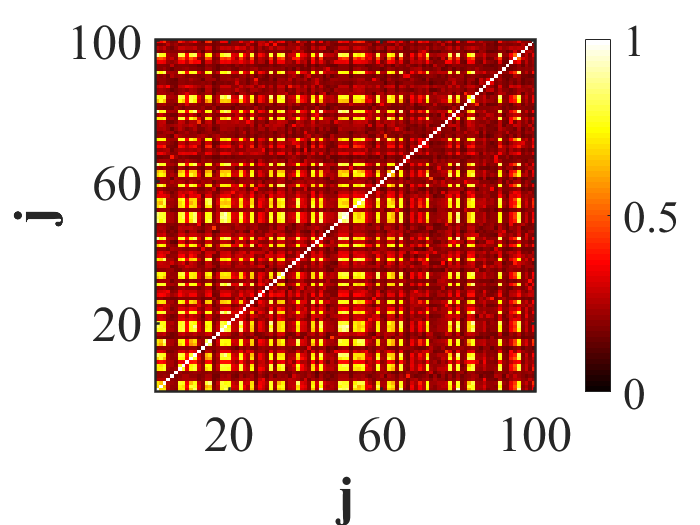}&
\includegraphics[width=4cm,height=4cm]{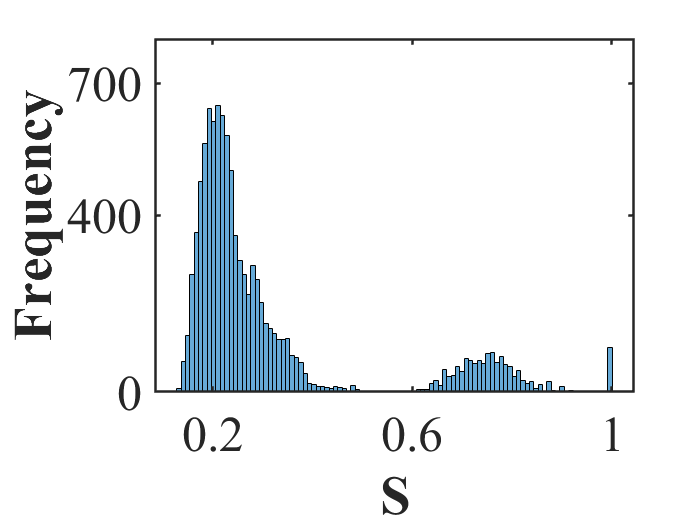}&
\includegraphics[width=4cm,height=4cm]{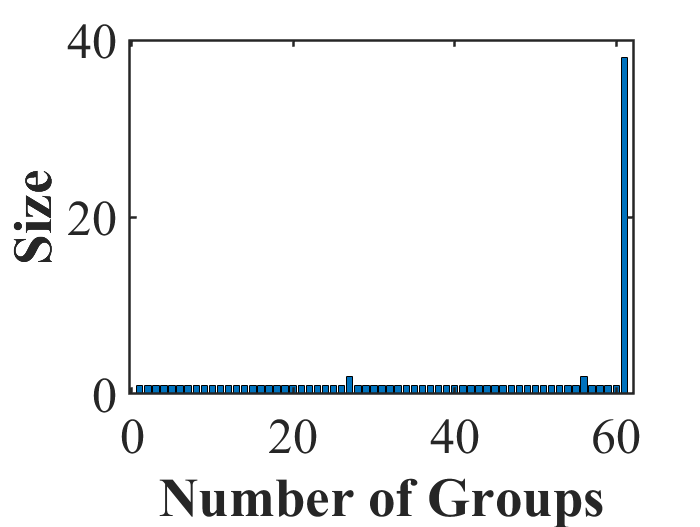}\\
(e)&(f)&(g)&(h)\\
\includegraphics[width=4cm,height=4cm]{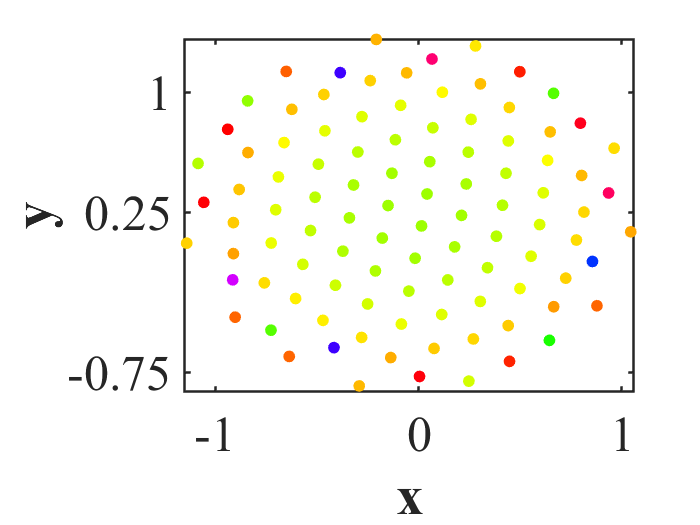}&
\includegraphics[width=4.5cm,height=4cm]{Newfigures/BS_100_S.png}&
\includegraphics[width=4cm,height=4cm]{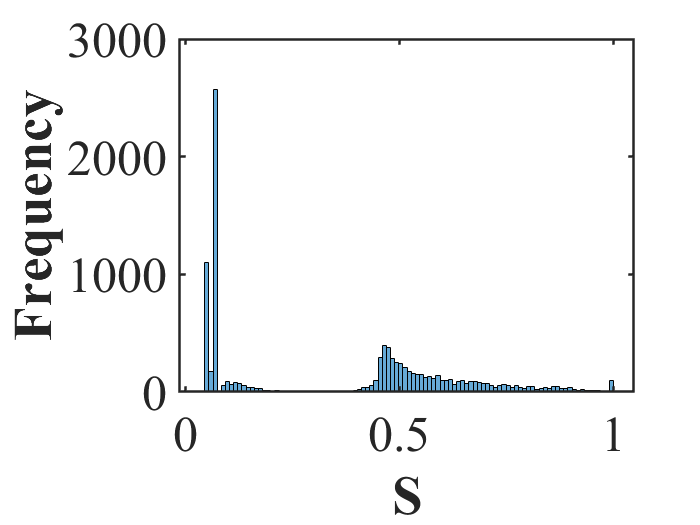}&
\includegraphics[width=4cm,height=4cm]{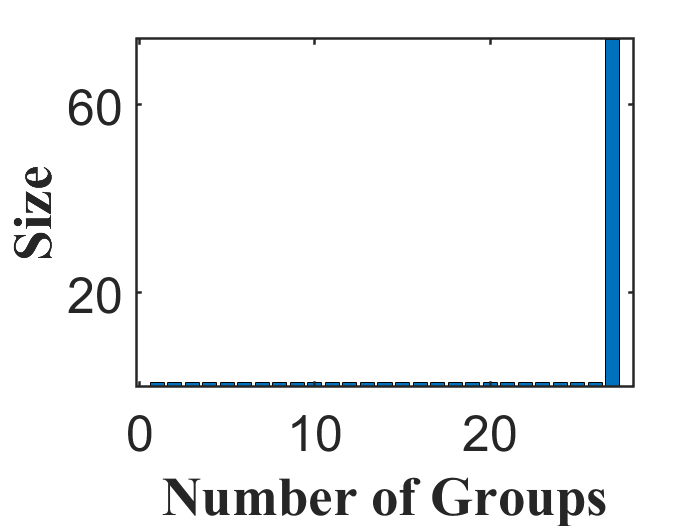}
	 \end{tabular}
	 \caption{For 100 nodes: (a)."Boiling state" (J,K,$\tau $)=(1,-0.85,3); and (b). boiling Chimera state (J,K,$\tau $)=(0.1,-1,2).Panels (a,e) show the spatial distributions of the oscillators. Panels (b,f) display the recurrence synchronization matrices (S), whose heterogeneous structures reveal the coexistence of oscillator pairs with different levels of dynamical correlation. Panels (c,g) present the corresponding distributions of (S), highlighting the presence of weakly, and highly correlated oscillators. Panels (d,h) show the group-size distributions extracted from (S); the coexistence of a dominant large group and several smaller groups indicates the simultaneous presence of coherent and incoherent populations, a characteristic feature of chimera-like states.}
	 \label{fig:jrpdensityboiling}
	 \end{figure*}

Analysis of the entropy of the Recurrence synchronization measure (S) shows high and fluctuating values, confirming the presence of unstable or intermittent structures, a generally chaotic but not random dynamics, marked by organized complexity.

By observing the histograms associated with the Recurrence synchronization measure matrices (S), we can see distinct peaks corresponding to areas of high and low coherence, respectively. Between these two extremes is an intermediate region, reflecting intermediate levels of correlation. To define the degree of independence (NL), we consider that this intermediate zone depends both on the system studied and on the simulation parameters chosen. Consequently, the independence threshold, which allows us to identify completely independent nodes, must be set between the two density peaks (in this intermediate zone). This adaptive approach ensures a more robust distinction between coherent, partially correlated and completely independent oscillators, and makes the indicator applicable to a variety of systems.

\begin{figure*}[htbp]
\centering
\includegraphics[width=19cm,height=9cm]{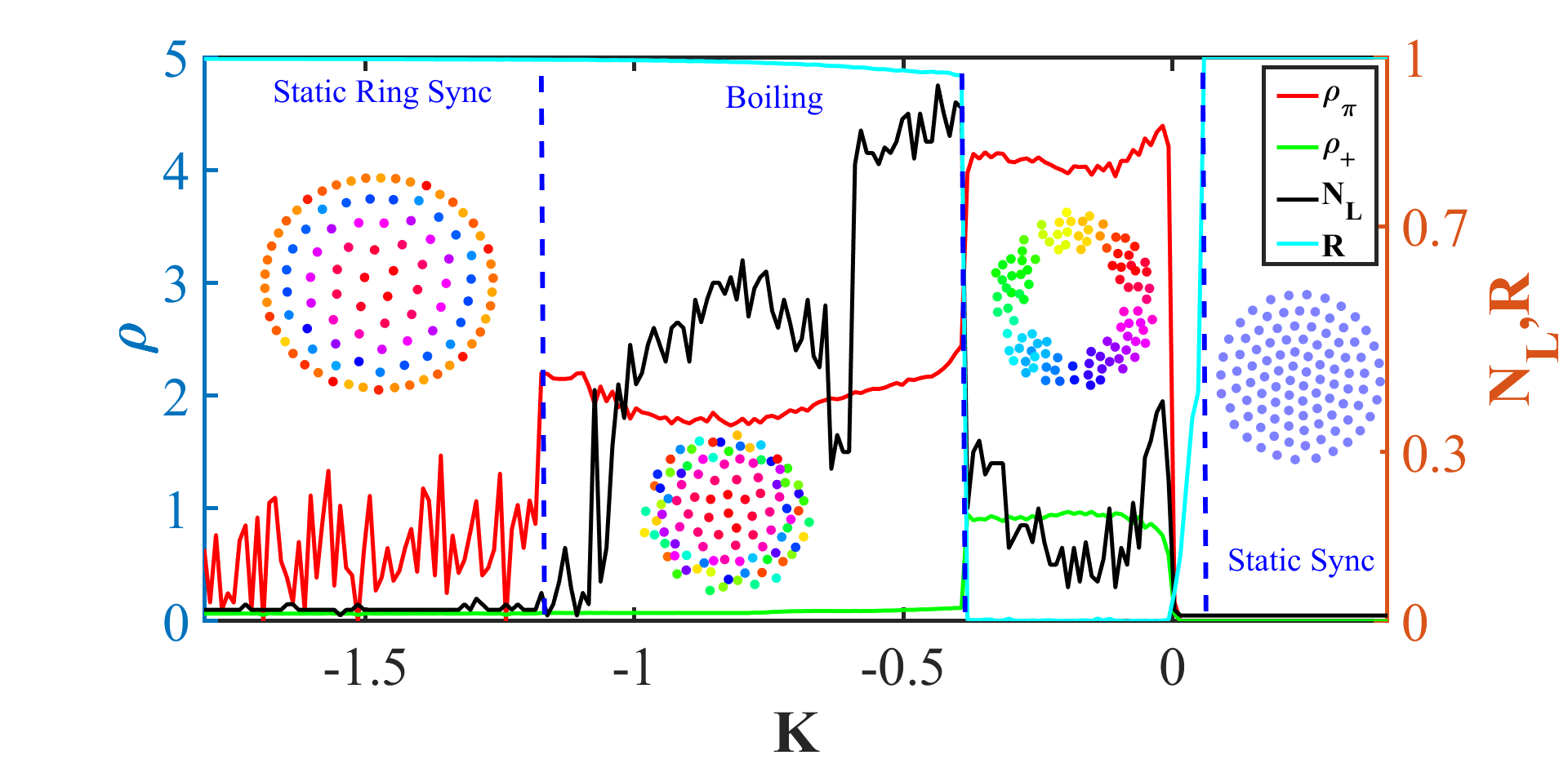}
\caption{Entropy, independence degree and order parameter function of phase coupling, detection of phase transition zones in swarmalators with delay (J,$\tau$)=(1,3). For fixed values of the parameters $J$ and $\tau$, the evolution of the coupling coefficient $ - 1.8 \le K \le 0.4$ reveals almost all of the dynamic states presented above. By gradually increasing $K$, the system successively passes through different collective regimes: from quasi-synchronization states to total synchronization, via the Boiling State,even Active Phase Wave and the Splintered Phase Wave for $ - 0.4 \le K \le 0$ . 
These transitions result in continuous and significant variations in recurrence measures, degree of independence, and coherence parameters, illustrating the richness of the system's dynamic behaviors and the sensitivity of the tools used to characterize them. 
} 
\label{fig:transit_swarm}
\end{figure*}


\section{Conclusion}\label{conclusion}

This work has made it possible to characterize different collective states, including chimera states, in a network of Colpitts oscillators and in a system of swarmalators with delay, based on recurrence analysis. We show that joint recurrence plots $(JRPs)$ and the histograms of Recurrence synchronization measure finely distinguish between complete synchronization, quasi-synchronization (static ring sync) and disorder regimes, whereas the classical order parameter gives very similar values for spatially different states. This increased sensitivity by the $JRP$ makes it possible to detect subtle phase shifts and quasi-coherence that cannot be detected by the order parameter. The Boiling State is confirmed as a chimera, marked by a coherent central region and an incoherent periphery.

The introduction of the degree of independence $(N_L)$ provides a direct quantification of the proportion of fully autonomous nodes, usefully complementing entropy($\rho$) and incoherence measures $(SI)$. These results confirm the relevance and potential of recurrence analysis for detecting and characterizing complex dynamic states, even in systems with variable topology, and open up prospects for the study of similar dynamics in neuroscience or distributed systems engineering.

\section*{Acknowledgments}
The author FFF and PL gratefully acknowledge the financial support by National Institute of Science and Technology in Innovative Research in Health Sciences – from Nanotechnology to Artificial Intelligence (INCT PICS) sponsored by Brazil’s National Council for Scientific and Technological Development (CNPq), grant no. 408417/2024-2 and grant no., Coordination of Superior Level Staff Improvement (Capes), grant no. 88887.197686/2025-00, and São Paulo Research Foundation (FAPESP), grant no. 2025/26818-7. Also, FFF and PL acknowledge the financial support by FAPESP grant no. 2025/18142-3, FFF thanks CNPq grants no.311989/2025-0. ADJ thanks capes grants no. 0001. PL thanks the support of the German Academic Exchange Service (DAAD) for funding his visit at the Potsdam Institute for Climate Impact Research (PIK) under the Grant Number (91897150). PL thanks Prof. Jürgen Kurths for fruitful discussions.

\nocite{*}
\bibliographystyle{prsty}

\end{document}